\documentclass[conference]{IEEEtran}
\IEEEoverridecommandlockouts
\usepackage{cite}
\usepackage{amsmath,amssymb,amsfonts}
\usepackage{algorithmic}
\usepackage{graphicx}
\usepackage{textcomp}
\usepackage{xcolor}
\def\BibTeX{{\rm B\kern-.05em{\sc i\kern-.025em b}\kern-.08em
    T\kern-.1667em\lower.7ex\hbox{E}\kern-.125emX}}

\begin{document}

\title{Aligning Robot Representations with Humans}

\author{\IEEEauthorblockN{Andreea Bobu}
\IEEEauthorblockA{\textit{University of California, Berkeley}\\
Berkeley, CA, USA \\
abobu@berkeley.edu}
\and
\IEEEauthorblockN{Andi Peng}
\IEEEauthorblockA{\textit{Massachusetts Institute of Technology} \\
Cambridge, MA, USA \\
andipeng@mit.edu}}

\maketitle

\begin{abstract}
As robots are increasingly deployed in real-world scenarios, a key question is how to best transfer knowledge learned in one environment to another, where shifting constraints and human preferences render adaptation challenging. A central challenge remains that often, it is difficult (perhaps even impossible) to capture the full complexity of the deployment environment, and therefore the desired tasks, at training time. Consequently, the \textit{representation}, or abstraction, of the tasks the human hopes for the robot to perform in one environment may be \textit{misaligned} with the representation of the tasks that the robot has learned in another. We postulate that because humans will be the ultimate evaluator of system success in the world, they are best suited to communicating the aspects of the tasks that matter to the robot. Our key insight is that effective learning from human input requires first explicitly learning good intermediate representations and then using those representations for solving downstream tasks. We highlight three areas where we can use this approach to build interactive systems and offer future directions of work to better create advanced collaborative robots.
\end{abstract}

\begin{IEEEkeywords}
Human-robot interaction, robot learning, representation learning.
\end{IEEEkeywords}

\section{Introduction}

\begin{figure*}[h]
\includegraphics[width=0.99\textwidth]{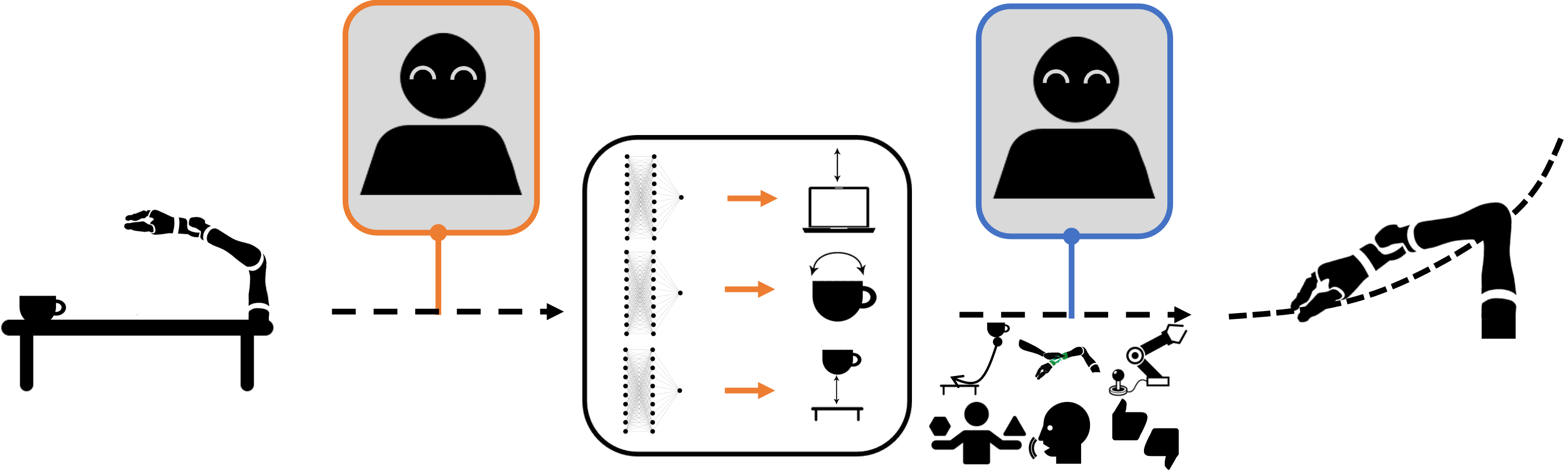}
\centering
\caption{Under our framework, the robot first learns \textit{human-guided representations} by asking the human for \color{orange}representation-specific input \color{black}to capture specific aspects of the task that they care about (e.g. distance to laptop, cup orientation, cup near table). The robot then uses the representation to learn how to perform the task from \color[HTML]{5072A7}task-specific input \color{black} like demonstrations, corrections, etc.}
\label{fig:front_fig}
\vspace{-4mm}
\end{figure*}

Imagine a world where you wake up in the morning, arise from bed, and your home robot assistant makes your bed. After getting ready, you head downstairs where your robot has placed a steaming mug of fresh coffee on the table exactly where it knows you will sit. After drinking the coffee, your robot picks up the empty mug and places it in the dishwasher as you leave the house and set off for work. The entire morning, your robot is incorporated seamlessly into your daily life and home. This scene of domestic bliss captures the essence of what we hope for from our advanced collaborative assistants -- the ability to effectively complete desired tasks while integrating into our environments and adapting to our individual preferences, akin to human-like collaboration.

%Today, researchers and engineers are able to directly teach autonomous systems how to perform advanced behaviors like these~\cite{abbeel2010autonomous,kolter2010probabilistic,wulfmeier2016maxentirl}.
Today, autonomous systems are increasingly able to learn advanced behaviors like those mentioned above~\cite{abbeel2010autonomous,kolter2010probabilistic,wulfmeier2016maxentirl}.
However, designing learning algorithms that match the adaptability and generalizability of human reasoning remains challenging: while these systems may perform their tasks successfully in the environment(s) and under the conditions they were trained on, their learned behaviors may not necessarily work well in novel deployment environments.
This problem can rear its head in a variety of instances: when physical constraints change (while it's okay for the robot to break mugs when trying out new grip poses in the lab, we may wish for them to be more careful in a home), when environment conditions, layouts, or compositions change (we may wish for the robot to grasp an octopus-shaped mug that it's never before seen), or when the task preferences of the human that the robot interacts with change (one human may prefer that the coffee is prepared as quickly as possible irrespective of mess, while another may prefer that the robot prioritizes not spilling the coffee while navigating the kitchen).

The key issue in all these cases is that, while the designer can anticipate some of the possible task specifications when training the robot, these specifications do not necessarily reflect the desires of the other humans the robot will interact with in its lifetime~\cite{ng1999policy,levine2020offline}.
In other words, the \textit{representation}, or abstraction, of the tasks the human hopes for the robot to perform in one environment may be \textit{misaligned} with the representation of the tasks that the robot has learned in another.
Our observation is that because humans have adapted their environments to capture the full idiosyncrasies of completing tasks that they desire, they are best equipped to help insert knowledge specifically describing aspects of the environment that are useful to the robot in the learning process. Specifically, human input can best help solve the \textit{representation alignment} problem of understanding what task aspects matter to the human when adapting to a new environment.
%In real-world deployments, a human will ultimately serve as the final evaluator of whether or not the robot has successfully accomplished the task.

Traditional methods of robot learning from human input instantiate representations as a set of hand-engineered \textit{features}---specific aspects of the task that a human may care about~\cite{ziebart2008maximum,HadfieldMenell2017InverseRD,abbeel2004apprenticeship,bajcsy2017phri,osa2018algorithmic}. These features are pre-specified by a system designer and function as state-space abstractions that insert structure for learning the task efficiently. However, they can be difficult to construct and impossible to exhaustively specify. Meanwhile, state-of-the-art deep learning methods~\cite{wulfmeier2016maxentirl,finn2016gcl,christiano2017preferences,fu2018learning,fu2018variational,brown2020brex,abbeel2004apprenticeship,torabi2018behavioral} bypass feature specification by operating directly on high-dimensional state spaces, thereby automatically constructing an \textit{implicit} representation from the person's task-specific input (e.g. demonstrations). Unfortunately, because these methods are optimized to learn the task while bypassing the explicit need to learn the representation, there is difficulty in disentangling the high-level representation from the specific task provided~\cite{fu2018learning,Reddy2020SQILIL,bobu2022inducing}. Consequently, effective task learning requires massive amounts of training data and renders generalization to new tasks difficult. In summary, one paradigm inserts useful structure to solve the robot learning problem efficiently but that structure is difficult to define; the other avoids explicitly specifying the structure but requires too much human data to extract it implicitly and thus struggles to generalize across different domains.

We postulate that effective learning from human input requires methodologies that combine the best of both traditional feature engineering and highly-expressive deep learning worlds. Our core idea is to \textbf{divide and conquer} the learning problem: \textit{explicitly} focus human input on teaching robots good intermediate representations before using those representations for downstream tasks. We call these \textit{human-guided representations}: abstractions that, if learned well, can enable robots to better solve tasks when deployed into the real world. We discuss several directions for learning human-guided representations as well as strategies for identifying misalignment and improving effective downstream task learning.
\section{Learning Human-Guided Representations}
\label{sec:HGR}

The representation learning literature has accrued a vast body of work on learning disentangled latent spaces in an unsupervised manner~\cite{chen2016infogan,Higgins2017betaVAELB,chen2018betaTCVAE}.
However, because these methods are purposefully designed to bypass direct human supervision, the disentangled factors in the learned embedding do not necessarily correspond to concepts in the human's representation. 
In other words, the robot's learned representation does not necessarily align with the human's, therefore adapting to how they want the task to be done is difficult.
%and this makes transfer from one environment to another difficult.
%This becomes problematic when we move between environments or tasks as it is difficult transfer the factors learned from one environment to another.
%We propose that because we are interested in aligning robots' task representations with humans', we will inevitably require a pipeline that allows for humans to guide learning.
Self-supervised learning inserts some human guidance by allowing for the designer to specify proxy tasks useful for feature learning~\cite{Doersch2015UnsupervisedVR,pathak2018zero,aytar2018playing,brown2020brex,laskin2020curl} (for example, predicting forward dynamics to capture what constrains movement). 
In this process, the human designer hopes to instill good representations into the robot by using their intuition to construct tasks which illustrate specific features.
%how the human prefers the coffee not be spilled onto the laptop.
However, devising proxy tasks is an exercise that requires nontrivial effort and expertise: human effort to manually specify features is instead traded for human effort to specify objective functions for extracting those features.
%Moreover, the designer hopes that these loss functions will implicitly lead to a representation that captures relevant features. But when the robot is moved into the real world, this representation may be still yet misaligned with the interfacing human. Ergo, similar to hand-crafting features, we must still yet manually construct new proxy tasks to adapt the robot to the new environment.

A more direct way to guide representation alignment is to learn directly from human input.
In standard imitation learning, the robot learns a policy that copies---or clones---human demonstrations ~\cite{osa2018algorithmic,abbeel2004apprenticeship}.
However, it cannot learn to imitate what it has not seen before, thus rendering human input non-generalizable to new tasks~\cite{levine2020offline,torabi2018behavioral}. Moreover, BC suffers from the problem of covariate shift, where once a learned policy drifts away from the demonstrations, errors compound more and more over time. 
Inverse reinforcement learning (IRL) attempts to extract a reward function from demonstrations that is intended to capture \textit{why} a specific behaviour is desirable~\cite{abbeel2004apprenticeship}, but unfortunately requires massive amounts of data to truly learn a fully-specified reward~\cite{fu2018learning,Reddy2020SQILIL}. IRL also requires expert or close to expert demonstrations \cite{ziebart2008maximum}.
Meta-learning reduces this sample complexity by reusing demonstrations from an array of different tasks in the training distribution~\cite{finn2017maml,xu2019metaIRL}, but ultimately still requires the human to know the test time task distribution \textit{a priori}, which brings us back to the manual specification problem: we now trade hand-crafting features for hand-crafting tasks.
%On the bright side, we can extract the feature representations automatically by aggregating demonstrations collected across the lifetime of the robot; on the other hand, it's unclear whether the space of possible tasks is smaller than the space of features that matter. \abnote{might want to make a bigger deal about this}

%Older IRL methods have looked at directly inferring a set of relevant features from task demonstrations, whether by projecting the state space to lower dimensions via PCA~\cite{vernaza2012efficient}, jointly inferring the reward and feature parameters~\cite{choi2013bayesian} or iterating between them by constructing features from a set of specified base features~\cite{levine2010feature, ratliff2007boosting}. These methods rely on engineering a relevant set of base features and underperform modern deep IRL methods which use more expressive architectures. \abnote{unsure what bigger picture to bring up with this paragraph, these are decade old methods...}

Because demonstrations are intended for teaching the robot \textit{how} to do the tasks, not \textit{what matters} for doing the tasks, they can only contribute to aligning representations implicitly. 
This might not result in learning algorithms extracting salient features that matter to the human for performing the desired tasks~\cite{bobu2022inducing}. 
As shown in Fig. \ref{fig:front_fig}, we propose that the robot should explicitly ask for \textit{representation-specific} human input to teach it the intermediate representation before using it to learn more generalizable downstream tasks from task-specific input.
%We call these \textit{intermediate human-guided representations}: abstractions that, if learned well, can enable robots to accomplish their tasks in a more generalizable way.
Importantly, because of this separation, these representations are not specific to any one particular task the human may want the robot to carry out; instead, they capture aspects causal for the potential task \textit{distribution} in the environment.

\textbf{Designing human input for representation learning.}
One option for learning intermediate human-guided representations is to instantiate them as feature sets like those in traditional methods, and let the human teach individual, novel features themselves~\cite{bobu2021ferl, bobu2022inducing}. 
%These features can then be composed into an intermediate representation that is more easily learnable by the robot.
A natural way to represent any specific new feature is via a neural network which is trained by asking the human for supervision labels representing the feature values at different states. Unfortunately, querying the human for labels to train this neural network requires a burdensome amount of human interaction. Even worse, humans are notoriously imprecise at giving these types of numerical inputs, rendering learned representations likely erroneous~\cite{Braziunas2008elicitation}. 
We propose that a key direction for future work is considering new types of representation-specific input that are highly informative about the feature without requiring too much effort from the human.
%We propose that a key direction moving forward is to enable the human to easily transform the learning problem into features that are more interpretable concepts and thus easier for them to teach. 
For example, a new type of structured human input called a \textit{feature trace}~\cite{bobu2021ferl}, where a human guides the robot from states where the feature is highly expressed to states where it is not, has been found to recover more robust and generalizable rewards with far less human effort. 
Moving forward, we can study additional forms of human input such as language or gaze and pose, that can also be targeted for feature learning.
Moreover, we can also consider types of human input that recover the feature representation as a whole (rather than one by one) via representation-specific proxy tasks -- \textit{calibration} tasks where the robot's goal is to specifically align itself with the demonstrating human.

\textbf{Transforming the representation for human input.}
Instead of designing the type of input the person can give to teach the representation, we can directly design the type of representation itself. 
Previously, when we instantiated the representation as a set of learnable features, we gave the human freedom to decide what feature each dimension of the representation was and provide feedback for teaching it to the robot. This enabled the human to add desirable task aspects to the representation even if the system designer did not originally think of them.
In some cases, though, it may be possible for the system designer to specify the necessary dimensions of the representation, just not the mapping to the representation itself.
This could happen, for example, if the designer has prior knowledge that the class of features the robot needs to express for its tasks has a well-studied representation.
%Instead of designing the type of input the person can give to teach the representation, we can design the type of representation. 
%Instead of designing structured types of representation-specific input the human can give to teach features efficiently, we can also turn our attention to the robot. Specifically, we consider how the robot can instantiate a representation for the learning problem as a latent space for which the human can provide more intuitive input.
%Instead of designing representation-specific human input, we can also turn our attention to the robot; specifically, to how the robot can instantiate a representation for the learning problem that the human can easily teach with regular input.
%As identified above, current methods which directly use raw demonstrations for task learning are impractical because of the data complexity required for diverse learning~\cite{levine2020offline,abbeel2004apprenticeship}. 
%If we can define a method to map intermediate (robot-understandable) representations into concepts that are designed specifically for human feedback, then we can leverage this mapping to transform natural human input into directly-learnable representations. 
For instance, recent work defines a model to relate emotions expressed in natural language, such as `happy' or `sad', into the Valence-Arousal-Dominance spectrum inspired by social psychology~\cite{sripathy2022teaching}. 
The human can teach the representation efficiently with natural language by having the robot map their utterances to their emotive latent VAD equivalent. 
This way, all user feedback for this representation contributes to learning about all emotions, and the robot can model new emotions that interpolate those seen during training. 
Moving forward, we should consider leveraging existing methods that define transformations of natural human-comprehensible concepts, such as language or images, into robot-comprehensible representations for downstream task learning~\cite{shridhar2022cliport,radford2021learning}.

\textbf{Designing the human-robot interface for learning.} In order to truly deploy collaborative robots in the world, we must eventually develop usable interactive interfaces that allow for effective information exchange of representations understood by both the human and robot. Existing work has highlighted the importance of the interface when a human and robot collectively share the same workspace, with key considerations being ease of use, specificity of communication, and reliability of feedback~\cite{wright2019agent,bansal2019updates}. Current methods suggest using visual displays, hand or face gestures, physical interaction and haptics, and verbal language can all be viable solutions towards effective human communication~\cite{berg2020review}. However, less work has been done in interfaces for how the robot can effectively communicate the representation of what it has learned with the human. For example, it would be desirable to have an interface by which the robot can effectively demonstrate or show the human what it \textit{thinks} is the correct desired task prior to actually deploying it in the real-world. This could be done in the form of mapping the proposed robot policy to simulated demonstrations or even natural language to communicate the intended behaviour. We propose that effective human-robot interaction which leads to learning human-guided representations will require the development of both streams of information flow in order to fully achieve its potential.

%\abnote{Might want to add another paragraph on how with HGRs we are essentially trying to recover features causal to the task \textit{distribution} in the environment, rather than to any particular task, like current LfD does. The idea is that upon seeing the environment, the person has a prior on the potential task distribution existent in that environment. What we aim to do for representation alignment is extract that from them.}
\section{Identifying Misalignment}

% build it up.
Along with learning transferable human-guided representations, it is also important to detect when misalignment exists in the first place. 
Misaligned representations may cause the robot to misinterpret the human's guidance for how to complete the task, execute unexpected or undesired behaviors, or degrade in overall performance~\cite{bobu2020quantifying}.
Ergo, we wish for the robot to \textit{know when it does not know} the aspects that matter to the human \textit{before} it starts incorrectly learning how to perform the task.
If misalignment is correctly detected, then a process which begins with expanding or re-learning the representation will better help ultimately learn the downstream task.
The key question is: how can the robot autonomously identify representation misalignment and know when to ask for help?

Several methods suggest an introspective approach where the robot can maintain uncertainty in its representation's ability to explain the human's input. 
By modeling humans as noisily rational agents choosing inputs in proportion to their exponentiated rewards~\cite{baker2007goal, jaynes1957infotheory, von1945theory}, Bayesian approaches can jointly infer both the reward parameter and a \textit{confidence} in whether the desired reward function can be captured by the current representation~\cite{fridovich-keil2019confidence,bobu2018learning,Losey2018IncludingUW, bobu2020quantifying,zurek2021situational}. When the human input refers to a reward that the robot's representation cannot support, the inferred confidence is low, signaling misalignment.
Meanwhile, deep learning methods often study this uncertainty through an \textit{ensemble} of neural networks~\cite{Lakshminarayanan2017ensemble,Sun2021OnCE}. The intuition here is that if multiple (identically trained) networks disagree on their predictions, this suggests that the input is out of distribution and therefore the learned representation is misaligned.

In both cases, once the robot detects misalignment there are a few options for how to proceed: discard the human input entirely, continue learning in proportion to its assessed confidence, or halt execution and ask the human to undergo the process of representation alignment from the previous section~\cite{bobu2020quantifying}.
Assuming the robot identified misalignment correctly, any of these options are viable alternatives to re-learning from the original human feedback.
Unfortunately, robustly detecting misalignment remains difficult in many real-world scenarios. We highlight three key areas where identifying misalignment is particularly challenging and offer brief suggestions for future work.

\textbf{Disambiguating between misalignment and noise.} 
When a robot's representation cannot explain the human input, it may be difficult to disambiguate whether this is due to representation misalignment or human noise~\cite{bobu2020quantifying}. This issue often arises from inexperienced users and is inherent to the types of data designers must work with in human-robot interaction scenarios.
A proposed, albeit expensive, method of addressing this challenge is to collect more data to balance out noise, but this solution would not fare well in online learning scenarios where the robot must detect misalignment in real time, from just a few observations.
We suggest that a more sustainable alternative is to investigate better human modeling for separating out these two sources of error~\cite{ramakrishnan2021bayesian}.

\textbf{Poor feature learning.} Misalignment can additionally occur due to two reasons: either the robot's representation does not fully capture an aspect that the human cares about or it does, but \textit{poorly}. The latter can occur if some of the features the robot learned were not learned well enough; for example, a feature might have required more data from the human in order to cover the state space and generalize to new areas. We propose that it is crucial for the robot to distinguish between misalignment due to an incomplete representation or due to incorrectly learned dimensions of the representation so that instead of attempting to re-learn a new feature, the robot knows to query for more data on the existing one. Future work is needed for understanding whether the robot needs to repair an existing learned feature, detecting which feature that might be, and developing interactive methods to elicit informative data to improve existing features.

\textbf{Feature confusion.} An even more fundamental issue exists when the human's input refers to something not captured by the robot's learned representation, but the representation nonetheless can explain their input. In this case, we have confused misalignment for human noise ~\cite{Sun2021OnCE,bobu2020quantifying}. This problem will especially occur if the representation is highly expressive and can only be solved by intaking additional human input: each input might be explainable by some hypothesis, but eventually no hypothesis can explain all input. More work is needed to study how to query for a broad and diverse set of human input, how the robot would best demonstrate the features it has learned to the human, and how to best balance between querying for data vs. learning with existing data.

\section{Learning the Downstream Task}

Once we have learned a human-guided representation, it is easy to then apply that representation towards learning a downstream task by using standard policy ~\cite{ng1999policy,levine2020offline,finn2017maml,levine2010feature} or reward learning techniques~\cite{jain2015learning,bajcsy2017phri,christiano2017preferences,brown2019extrapolating,fu2018variational,HadfieldMenell2017InverseRD,shah2018state}.
However, human-guided representations have important implications for how they impact the downstream learning pipeline. We subsequently discuss three considerations that future work should consider to fully close the learning loop. 

\textbf{Using the right features at the right time.}
In this proposal, we have advocated for learning a human-guided representation that is sufficiently decoupled from any specific task the human may have provided feedback for and focuses instead on capturing causal aspects for the potential task distribution in the environment. 
When the robot specializes on a task, the representation by construction will contain features that are irrelevant for that task. If all feature dimensions in the representation were orthogonal to one another, this would not cause any issue. However, in the real world, many relevant features may be related and, thus, \textit{spurious correlation} between features could affect task learning~\cite{dehaan2019causal}. 
Future directions of work should enable the robot to \textit{focus on the right features at the right time}. One idea for accomplishing this is to employ feature selection strategies to activate the subset of the representation that matters for the specific task at hand. This strategy could be heuristic-based, like choosing the minimum set that maximizes coverage~\cite{sax2018midlevel}.
%Work in computer vision has shown that selecting subsets among many types of visual features can improve sample complexity for learning downstream tasks~\cite{sax2018midlevel}.
%We similarly suggest that when specializing, the robot should additionally have a feature selection or attention mechanism to activate the subset of the representation that matters for the specific task at hand. 
Alternatively, since we would hope for our learned representations to be more human interpretable in nature, we could also consider building interfaces where the person themselves can quickly indicate to the robot which features are important for the specific task they want~\cite{cakmak2012questions}.

\textbf{Using representations to better understand humans.}
Human-guided representations also enable us to learn something about how the person generates the task input in the first place. In particular, the previously mentioned human decision-making models~\cite{baker2007goal,von1945theory,Luce1959choice} assumed that, out of a set of choices, the person selects their input in proportion to these choices' exponentiated rewards. However, we suggest that human-guided representations inform the robot how it should interpret the person's task input, thus we should \textit{reinterpret the available choices from the perspective of the learned representation}~\cite{bobu2020less}. We suggest future research must revisit how popular robot learning methods are affected by reinterpreting human input through the lens of their representation.

\textbf{Grounding representations to real-world tasks.}
Much of HRI has historically assumed that the robot already has access to all the aspects in the environment that the interacting human might care about. This assumption has enabled researchers to make progress on human-robot collaborative algorithms without needing to worry about how to formally ground the robot's behaviour to complex environments and tasks that we would see in the deployment scenarios.
Human-guided representations can help bridge the gap towards learning from high-dimensional state spaces as we know the real-world to be, opening the door to HRI applications more challenging and tractable than ever before.
\section{Conclusion} 
Ultimately, the true evaluators of any system deployed in the real world will be the humans that it interacts with, and thus soliciting input from them to effectively learn downstream tasks appears critical. Learning effective methods to learn from human input holds the promise of enabling more advanced, collaborative human-aligned robotic systems. In this paper, we proposed several methods for learning more generalizable intermediate representations from humans and suggested directions for moving towards a more continual and interactive learning framework. It is through understanding and utilizing this bi-directional communication flow that truly effective human-robot collaboration can exist.

%\section*{Acknowledgment}

{
\bibliographystyle{IEEEtran}
\bibliography{IEEEabrv,references}
}

\end{document}